# Integrated Warehouse Location and Inventory Decisions in a Multi-location Newsvendor Problem

Jianing Zhi

School of Business

Jiaxing University

Guangqiong Rd No.899, Jiaxing, Zhejiang Province, China, 314001

email: zhijianing@zjxu.edu.cn

Xinghua Li

College of Control Science and Engineering

Zhejiang University

38 Zheda Road, Hangzhou, Zhejiang Province, China, 310027

email: lixinghua0620@zju.edu.cn

Zidong Chen

School of Computer Sciences

Universiti Sains Malaysia

Gelugor, Penang, Malaysia, 11800

Email: chenzidong@student.usm.my

# Abstract

In this paper, we investigate a supply chain network with a supplier and multiple retailers. The supplier can either take orders from retailers directly, or choose to build a warehouse somewhere in the network to centralize the ordering from retailers. Meanwhile, we take three modes of transportation cost into account, including distance-dependent, quantity- dependent, and distance-quantity-dependent costs, to formulate six models. For the three decentralized models, we provide closed-form solutions to compute the optimal order quantity of each retailer. For the centralized models, we develop closed-form solutions for the first two models as the transportation cost only depends on either distance or order quantity; but when it depends on both, the model becomes a non-linear programming problem. We develop a solution algorithm named Q-search to find a high-quality solution that includes the order quantity and the warehouse location. The experiment results show that the centralized model outperforms the decentralized model in large networks in terms of profit and service level.

# 1 Introduction

The classical newsvendor problem is reflective of many real-life situations, and it is widely applicable for decision making in various areas such as fashion, sporting goods, and electronics industries, at both manufacturing and retail levels. The classical newsvendor problem determines the order quantity for a single product during a single period by maximizing the expected profit under probabilistic demand (Silver et al., 1998). The newsvendor model considers a shortage cost associated with each demand that cannot be fulfilled. Also, if there is any stock remaining in inventory at the end of the period, it is sold at a salvage value, or it is disposed of. Therefore, there is a trade-off between ordering/producing too little and too much due to the uncertainty of the demand.

The classical newsvendor problem may not apply to some modern businesses since currently many companies tend to adopt new expansion trends to make shopping more convenient and to elevate market share. To address this practical need, researchers have studied the multi-location newsvendor problem. Khouja (1999) provides an extensive overview that summarizes the variants of newsvendor problem. In this paper, we consider a multi-location newsvendor problem where the demand at each location is satisfied by the same supplier. More specifically, our model consists of a single supplier and multiple newsvendor locations, each of which observes a random demand for a seasonal (short-life time) product. In the traditional (decentralized) system, before the sales period, each retailer (newsvendor) places an order to meet the random demand, which is assumed to be independent of the demand of other newsvendors. The retailer determines the order quantity in a way such that it maximizes its profits. On the other hand, upon receiving the orders, the supplier ships the order to each retailer in a timely manner. Therefore, each party (the supplier and the retailers) in the supply chain will maximize its gain. We refer to this system as the decentralized system (DS), as shown in Figure 1a.

Many researchers studied the multi-location newsvendor problem and noted that when demands from several retailers are consolidated, the expected holding and penalty costs of the consolidated system are lower than that of the decentralized system (Chang and Lin, 1991; Cherikh, 2000; Eppen, 1979; Lin et al., 2001; Yang et al., 2021; Govindarajan et al., 2021). To take advantage of the centralized ordering and the economies of scale in transportation, the supplier may prefer each retailer to order through a distribution center (DC). Hence, the second supply chain system we consider is a centralized system where a DC exists to coordinate and consolidate the orders from the retailers. In the centralized system, the supplier oversees the whole supply chain and fulfills demands of retailers through the DC, as shown in Figure 1b. The DC receives a single consolidated shipment from the supplier and distributes the orders to the retailers when demand occurs. Many large retail industries and car manufacturers opt for large warehouses that distribute orders to local stores and dealerships across a region, rather than carrying separate inventory at different locations. Moreover, the idea of centralizing stocks has been widely adopted in the fast-growing e-commerce industry.

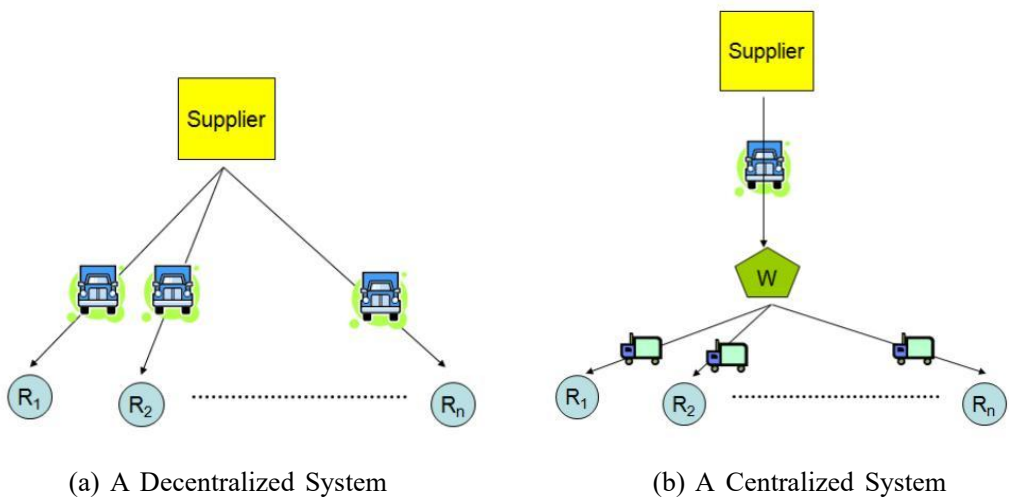

(a) A Decentralized System  (b) A Centralized System

Figure 1: Decentralized and centralized systems

The motivation for these models stems from recent supply chain management (SCM) practices. First of all, in the last decade, SCM strategies that transfer stock level decisions and stock-keeping responsibilities from downstream parties to the upstream have become popular. These new approaches include vendor managed inventory (VMI) (Cachon, 2001; Cetinkaya and Lee, 2000; Cetinkaya et al., 2006; Toptal and Cetinkaya, 2006), consignment vendor managed inventory, vendor hub (Lee and Chu, 2005), drop-shipping (Netessine and Rudi, 2004), and Click-and-Mortar (Lee and Whang, 2000). Under the VMI process, retailers can share demand information with a supplier who should then make the stock replenishment decisions for both itself and the retailers. A variant of VMI is consignment VMI, in which the supplier makes the stock replenishment decisions and owns the stock until it is sold. Currently, Walmart, one of the leading retailers, operates under a consignment VMI process. Vendor hub is an another popular method adopted by numerous electronics/computer manufacturers. These manufacturers request a supplier to keep a minimum stock in a hub, and they only pay the supplier when goods are pulled and consumed. The increasing popularity of these models drives us to wonder whether the gains obtained from the centralized system are superior to the ones from the decentralized system under different settings of transportation cost.

Another motivation comes from the utilization of transportation. In the decentralized system, the supplier ships the orders separately to each retailer and possibly over long distances. Moreover, these trucks may or may not be full depending on the order size, causing a potential waste which could be reduced by consolidation. For this purpose, the supplier may choose to outsource transportation to a third party logistics provider. Hence, explicit modeling of transportation costs for both decentralized and centralized systems not only provides insights for the supply chain managers, but also for the third party logistics companies.

The questions we aim to answer include:

- How do decentralized and centralized solutions differ?
- What would be the payoffs for each party in the supply chain? and if there are any savings, how would these savings be distributed in the supply chain?
- Under what circumstances it is preferred to implement a centralized system?

- How does the DC location impact the results?

- How do different transportation costs impact the results? In other words, how would the results change given economies of scale and/or economies of distance?

Although there have been studies on the impact of inventory centralization, no prior research has explicitly modeled the relation between inventory ordering and the DC location in a multi-location newsvendor problem. Moreover, the explicit modeling of the transportation cost provides further insights for the relation between the centralized and decentralized systems. We summarize the contributions of this work as follows:

- We develop two newsvendor models, including a decentralized system model (DSM) and a centralized system model (CSM) under a practical newsvendor setting. Both models help decision makers of a firm maximize the overall profit with estimated demand following a normal distribution. Besides, CSM determines the optimal DC location where a DC is suggested to build.

- We compare DSM and CSM regarding three modes of transportation cost, and develop solution approaches to these particular problems. In addition, we consider a change on CSM: instead of setting a DC at the optimal location, it may be cost-efficient to set it at the closest retailer to the optimal location. This option is referred to as Retailer-as-DC, and presents a high practical value.

- We conduct a suite of experiments to validate our models and solution approaches. The results demonstrate that 1) CSM outperforms DSM in large-scale network with more retailers, 2) the proposed models work well for simulated demands, with a result in line with the one obtained from expected demands, and 3) Retailer-as-DC shows great profitability and is highly recommended if the chosen retailer is sufficiently close to the optimal DC location.

The rest of this paper is organized as follows. Section 2 provides a review of related work. Section 3 presents the assumptions and notations. In Section 4, we formulate the DSM and CSM models and develop solution approaches for each model. Section 5 is a comparison of DSM and CSM. We validate our models and solution approaches in Section 6, and conclude the paper in Section 7.

# 2 Literature Review

We provide a review of related studies in two areas: multi-location newsvendor problems and location-inventory problems.

## 2.1 Multi-Location Newsvendor Problem

The single period newsvendor problem has been widely discussed in the literature. Khouja (1999) and Silver et al. (1998) provide comprehensive reviews on this subject. Eppen (1979) proposes a multi-location newsvendor problem with normal demand and linear holding and penalty cost functions at each location. This paper shows that the total cost in a decentralized model exceeds the one in a centralized system with a difference that depends on the correlation of demands. For uncorrelated and identically distributed demands, however, the expected cost of a centralized facility increases as the square root of the number of consolidated centers. Chang and Lin (1991) extend Eppen's model by taking transshipment cost among multiple retailers into account. The authors give the condition of cost savings on a centralized model compared with a decentralized model with certain functions of holding cost, penalty cost and transshipment cost, in which the transportation cost only includes the transshipment of excess order quantities among retailers but ignores the transportation from a supplier to retailers. Slikker et al. (2005) also consider a multi-newsvendor problem with transshipment among retailers. Different from Chang and Lin (1991), the authors analyze the problem with a cooperative game and show that stable divisions of expected profits also exist if transshipment costs are positive. Furthermore, they show that given different retail and wholesale prices rather than anonymous prices, stable divisions of expected profits exist as well. Cherikh (2000) also recommends centralization with a condition that a portion or all of the excess demand at a location can be reallocated among other newsvendors. Lin et al. (2001) develop a single-period, single-product inventory model with several individual sources of demand. They prove that the Eppen's property still holds under a multi-source demand setting. Hartman and Dror (2005) consider this problem with normally distributed and correlated individual demand. They develop a greedy heuristic procedure to find the optimal centralized policy. Lee and Chu (2005) discuss whether it is profitable to move the stock-keeping responsibility to the upstream (supplier) in a newsvendor inventory control system. They find that the critical fractile used in a newsvendor problem to determine the optimal stock level can often be used as a yardstick to quickly assess the merits of adopting the new mechanism. Furthermore, they show that when the critical fractile of the supplier is greater than that of retailers, adopting the new mechanism always makes both members better off, provided that some risk-sharing rules can be adopted. Ji and Shao (2006) consider a model with fuzzy demands and quantity discounts in a hierarchical decision system, in which the supplier decides the wholesale prices of newspaper to maximize its profit, while the retailers decide the order size of newspaper to maximize their expected profits. Belgasmi et al. (2013) proposes a more realistic multiobjective study framework for multi - location Transshipment and Newsvendor inventory models, where it considers the aggregate cost, fill rate, and shared inventory quantity as conflicting objectives and employs two reference multiobjective evolutionary algorithms (SPEA2 and NSGA - II) for solution, while taking into account the presence of storage capacity constraints. Fard et al. (2019) introduces a continuous-time Newsvendor model, which considers non-risk-neutrality, uses demand-price correlation to choose optimal order quantity and financial portfolio for utility maximization, and shows it needs less inventory buffer to handle demand uncertainty than myopic methods, thus better utilizing capital. Xiao and Wang (2023) investigate a multi-location newsvendor model incorporating either additive or multiplicative random yield assumptions. Subsequently, they derive the expected cost and variance of cost under both centralized and decentralized systems.

Although researchers have extended the multi-location newsvendor problem in various di- rections, the transportation cost has not been thoroughly considered, and the characteristics of the centralized DC and transportation modes have not discussed. Our research attempts to fill this gap and provide inventory-location joint models to analyze how the location and the capacity of a DC influence a centralized newsvendor system.

## 2.2 Location-Inventory Problem

In general, the location-inventory problems have two types of decisions: the DC location and the optimal order quantity of each retailer. In most prior studies, the locations of DCs are selected from a finite set of candidate locations, which makes a discrete location model. Our model is a continuous problem, as it decides the coordinates of a DC instead of choosing it from candidate locations. For the inventory decisions, many prior studies focus on periodic- review inventory policies such as the (r, q) policy (insert reference) or the approximation of this policy, which is essentially the classic economic order quantity (EOQ) model. It is reasonable to combine the long-term decision on a strategic location with the periodic inventory context, since both decisions are made for a finite long period. In our problem, we consider the location problem with a single period newsvendor setting, as the demand for seasonal goods in a selling season may boost quickly and drop sharply after the critical selling period. For example, the demand for eastern eggs increases dramatically in April, while there is no demand at all for the rest of a year. Retailers would like to increase the inventory of eastern eggs temperately and do not want it to influence the regular stock of other products. Thus, a temporary warehouse is a reasonable choice to hold the seasonal inventories, which motivates us to combine the location problem with a newsvendor structure.

Shen et al. (2003) propose a joint location-inventory model in which location, shipment, and nonlinear safety stock inventory costs are taken into account in the same model. They develop an integrated approach to determine the number of DCs to establish, the DC locations, and the magnitude of inventory to maintain at each DC. In their model, the potential DCs are built at existing retailers rather than setting new facilities as central DCs.

Tancrez et al. (2012) study a location-inventory problem in a three-level supply network. The authors formulate a nonlinear model including transportation and fixed holding costs, which decomposes into a closed-form equation and a linear program with fixed DC flows. They develop an iterative heuristic procedure that estimates the DC flows a priori, solves the linear program, and then improves the DC flow estimations.

Diabat et al. (2015) address a closed-loop location-inventory problem involving a single-echelon forward supply chain (DCs distributing a product to retailers with random demand) and a single-echelon reverse supply chain (RCs remanufacturing returns into spare parts and redistributing them). The objective is to select optimal DC and RC locations and assign retailers to them. The problem is modeled using a mixed integer nonlinear approach and solved via a two-phase Lagrangian relaxation algorithm.

Puga and Tanvrez (2017) tackle a location–inventory problem for large supply chain networks with uncertain demand, introducing a continuous non-linear model integrating key decisions and costs. They propose a heuristic algorithm to efficiently solve the linearized problem iteratively, with computational experiments showing its effectiveness in finding near-optimal solutions quickly. Managerial insights are also provided on the impact of demand uncertainty, risk pooling, and retailer safety stocks on supply chain design.

Song and Wu (2023) delve into two pivotal problems: the Location-Inventory Problem (LIP) and the Location-Inventory-Routing Problem (LIRP). A Mixed Integer Nonlinear Programming (MINLP) model is crafted for LIP, while Mixed Integer Linear Programming (MILP) models are put forth for LIRP. Subsequently, the developed mathematical models and solution methodologies are applied to a real-world supply chain network for fresh milk.

Li et al. (2023) address the location–inventory problem, optimizing distribution center (DC) location, inventory, and allocation for multiple retailers with uncertain demand. Unlike prior studies assuming linear or concave facility operating costs, they model it as an inverse S-shaped function, reflecting economies and diseconomies of scale. Then, they formulate the problem as a set-covering model and solve its linear relaxation using a column generation (CG) algorithm, with a branch-and-bound (B&B) method efficiently tackling the pricing problem.

# 3 General Assumptions and Notation

We consider a supplier that markets and delivers goods in a single selling season to n retailers. This setting fits the supply chain for many seasonal products such as fashion goods, holiday items, and fast-moving electronic products. We assume that in both DSM and CSM, the locations of the supplier and the geographically dispersed retailers are known. The DC location in the CSM is continuous and needs to be decided. The selling season is relatively short and has a well-defined beginning and end. Therefore, retailers must decide how much to order before the start of the season. Each retailer faces a random demand, $x$, that is determined by a continuously differentiable probability density function $f_i(x)$ with a support on $[0,\infty)$. The cumulative distribution function (CDF) of $f_i(x)$ is denoted as $F_i(x)$. Every retailer experiences stochastic demand which is unknown at the moment of ordering.

In the DSM, retailers receive orders directly from the supplier, meaning that the supplier needs to solve n newsvendor problems with transportation cost for all retailers. On the contrary, in the centralized model, the supplier observes the demand distribution of all retailers and then determines a total order quantity $Q_0$ shipped to the DC. Then, at the beginning of a selling season, after observing the actual demand at retailers, the DC will send the actual order $x_i$ to each retailer. If any inventory remains at the end of the period, a discount is used to sell it, or it is disposed of (i.e., salvage value). We assume that since there is only one kind of product, and the salvage value is the same for all retailers. If the order quantity at a retailer is lower than the actual demand at that retailer, a shortage cost occurs to penalize the lost profit.

Transportation cost is a critical component in our model. In the DSM, there is a transportation cost from the supplier to retailers. In the CSM, there are transportation costs between the supplier and the DC and also between the DC and retailers. Our model considers the transportation cost as a function of traveled distance and/or order size.

We define the following notations:

**Parameters:**

| | |
|---|---|
| $\mathcal{I}$ | set of retailers, $i = 1, \ldots, n$. |
| $L_0$ | location of the supplier, $L_0 = (a_0, b_0)$. |
| $L_i$ | location of retailer $i$, $L_i = (a_i, b_i)$, $\forall i \in \mathcal{I}$. |
| $x_i$ | random quantity demanded at retailer $i$, $\forall i \in \mathcal{I}$. |
| $\mu_i$ | the mean of demand at retailer $i$, $\forall i \in \mathcal{I}$. |
| $\sigma_i$ | the standard deviation of demand at retailer $i$, $\forall i \in \mathcal{I}$. |
| $f_i(x_i)$ | the probability density function of $x_i$ with support on $[0, \infty)$. |
| $F_i(x_i)$ | the CDF of $x_i$. |
| $x_0$ | total demand at DC, $x_0 = \sum_{i \in \mathcal{I}} x_i$. |
| $f_0(x_0)$ | the density function of the total demand $x_0$ with support on $[0, \infty)$. |
| $F_0(x_0)$ | the CDF of $x_0$. |
| $s$ | selling (retail) price per unit, $\forall i \in \mathcal{I}$. |
| $c$ | unit cost of supplier. |
| $v$ | salvage value per unit. |
| | It is assumed that $s > c > v$. |
| $b$ | shortage (out-of-stock) cost per unit at retailer $i$, $\forall i \in \mathcal{I}$. |
| $K_i$ | retailer's fixed cost of replenishment, $\forall i \in \mathcal{I}$. |
| $K_0$ | fixed replenishment cost for the supplier. |
| $ds_i$ | distance between the supplier at $L_0$ and retailer at $L_i$. |
| $TS_i(Q_i, ds_i)$ | transportation cost between the supplier/DC and retailer $i$, $\forall i \in \mathcal{I}$. |
| $T0(Q_0, d_0)$ | transportation cost between the supplier and the DC. |

**Decision variables:**

$Q_i$      order size of retailer i from the warehouse, $\forall i \in \mathcal{I}$.

$Q_0$      order size of the distribution center from the supplier.

$X$      location of the DC, $X = (x, y)$.

$d_0$      distance between the supplier at $L_0$ and the DC at $X$.

$d_i$      distance between the DC at $X$ and retailer at $L_i$.

$\mathcal{D}$      $\{d_0, d_1, \ldots, d_n\}$

# 4 Models and Solutions

In this section, we formulate DSM and CSM. For each model, we consider three modes of transportation costs, which influence the problem complexity. Moreover, we develop the solution approach for each proposed model.

## 4.1 Model 1: Decentralized System Model (DSM)

We can decompose the DSM into multiple independent small problems, each of which consists of the suppler and a retailer and can be optimized individually. In particular, the problem at retailer $i$ is to determine the order quantity $Q_i > 0$ by maximizing the expected profit $\Pi_i(Q_i) = E[\pi_i(Q_i)]$, where $\forall i \in \mathcal{I}$, $\pi_i(Q_i)$ is:

$$\pi_i(Q_i) = \begin{cases} sx_i - wQ_i - b(x_i - Q_i) - K_i - TS_i(Q_i, ds_i) & \text{if } x_i \geq Q_i \\ sx_i + v(Q_i - x_i) - wQ_i - K_i - TS_i(Q_i, ds_i) & \text{if } x_i < Q_i \end{cases}$$

On the other hand, the supplier solves a problem to maximize its expected profit, $\Pi_{sp}(Q_1, Q_2, \ldots, Q_n)$:

$$\Pi_{sp}(Q_1, Q_2, \ldots, Q_n) = \sum_{i \in \mathcal{I}} (w - c)Q_i - K_s$$

Therefore, the total profit obtained in DSM is

$$\Pi^D(Q_1,\ldots,Q_n) = \Pi_{sp}(Q_1,Q_2,\ldots,Q_n) + \sum_{i \in \mathcal{I}} \Pi_i(Q_i)$$

$$= \sum_{i \in \mathcal{I}} (s\int_0^\infty x_i f(x_i)dx_i - b\int_{Q_i}^\infty (x_i - Q_i)f(x_i)dx_i$$

$$+ v\int_0^{Q_i} (Q_i - x_i)f(x_i)dx_i - cQ_i - TS_i(Q_i, ds_i) - K_i) - K_0$$

subject to $\quad F_i(Q_i) \geq \gamma, \forall i \in \mathcal{I}$

The transportation cost we consider in this model has both fixed and variable parts. We assume that transportation costs follow three different structures. Let $T_i(Q_i, d_i)$ be the transportation cost between the supplier and retailer $i$; meanwhile, we define $T_i$ as a function of quantity $Q_i$ and/or distance $d_i$. There are three cases.

**As a function of quantity shipped:**

$$TS(Q_i, ds_i) = ps_i^q + rs_i^q Q_i \quad (DSM)$$
$$T0(Q_0, d_0) = p_0^q + r_0^q Q_0 \quad (CSM)$$
$$TI(Q_i, d_i) = p_i^q + r_i^q Q_i \quad (CSM)$$

**As a function of distance traveled:**

$$TS(Q_i, ds_i) = ps_i^d + rs_i^d ds_i \quad (DSM)$$
$$T0(Q_0, d_0) = p_0^d + r_0^d d_0 \quad (CSM)$$
$$TI(Q_i, d_i) = p_i^d + r_i^d d_i \quad (CSM)$$

**As a function of quantity and distance:**

$$TS(Q_i, ds_i) = ps_i^{qd} + rs_i^{qd} Q_i ds_i \quad (DSM)$$
$$T0(Q_0, d_0) = p_0^{qd} + r_0^{qd} Q_0 d_0 \quad (CSM)$$
$$TI(Q_i, d_i) = p_i^{qd} + r_i^{qd} Q_i d_i \quad (CSM)$$

in which $ps_i^q$, $ps_i^{qd}$ and $ps_i^d$ are the bundling cost that is associated with one shipment. Meanwhile, $rs_i^q$ is transportation cost per unit product, $rs_i^{qd}$ is the transportation cost per mile per unit product, and $rs_i^d$ is the transportation cost per mile.

## 4.2 Solution of DSM

**Case 1:** $TS_i(Q_i, ds_i) = ps_i^q + rs_i^q Q_i$

When the transportation cost has a linear relation with the order quantity, we can consider the unit transportation cost $\alpha^q$ as a part of the unit variable cost. Therefore the optimal order quantity of each retailer $i$ has the traditional newsvendor problem solution structure. For each retailer we have:

$$\frac{d}{dQ_i}\Pi^{D1}(Q_i) = s\mu_i - cQ_i - bE[x_i - Q_i]^+ + vE[Q_i - x_i]^+$$
$$= (s-c)\mu_i - (b-c)E[x_i - Q_i]^+ - (c-v)E[Q_i - x_i]^+$$
$$= (s-c)\mu_i - G(Q_i)$$

Where

$$G(Q_i) = (b-c)E[x_i - Q_i]^+ + (c-v)E[Q_i - x_i]^+$$

Then maximizing the total profit is equivalent to minimizing $\sum_{i \in \mathcal{I}} G(Q_i)$. The optimal order quantity of each retailer is:

$$Q_i^* = F_i^{-1}(\frac{c_u}{c_u + c_o}), \forall i$$

where $c_u = b - c - rs_i^q$, $c_o = c + rs_i^q - v$, and $F_i^{-1}$ is the CDF of $x_i$ following $(\mu_i, \sigma_i)$. The second derivative is $\frac{d^2}{dQ_i}\Pi^{D1}(Q_i) = -(b-v)f(Q_i) \leq 0$, implying a concave profit function is concave on $Q_i$. Also, since the constraint set $F(Q_i) \leq \gamma$ applies, we have the following solutions of DSM when the transportation cost is linearly dependent on $Q_i$:

$$Q_i^* = \max\{F_i^{-1}(\gamma), F_i^{-1}(\frac{b-c-rs_i^q}{b-v})\}$$

**Case 2:** $TS_i(Q_i, ds_i) = ps_i^d + rs_i^d ds_i$

When the transportation cost is linearly dependent on the distance between the two end points, the cost function is irrelevant to order size $Q_i$. Therefore we can solve the simple newsvendor model to obtain the optimal solution.

$$Q_i^* = F_i^{-1}(\frac{c_u}{c_u + c_o}), \quad \forall i \in \mathcal{I}$$

where $c_u = b - c$, $c_o = b - v$. Also,

$$Q_i^* = \max\{F_i^{-1}(\gamma), F_i^{-1}(\frac{b-c}{b-v})\}, \quad \forall i \in \mathcal{I}$$

**Case 3:** $TS_i(Q_i, d_i) = ps_i^{qd} + rs_i^{qd} Q_i ds_i$

When the transportation cost is dependent on both order quantity and distance, we still have similar result as case 1 and 2, because in DSM locations of the supplier and retailers are given.

$$Q_i^* = F^{-1}(\frac{c_u}{c_u + c_o}), \forall i \in \mathcal{I}$$

where $c_u = b - c - rs_i^{qd}$, $c_o = c + rs_i^{qd} - v$. We can rewrite $Q_i^*$ as:

$$Q_i^* = \max\{F_i^{-1}(\gamma), F_i^{-1}(\frac{b-c-rs_i^{qd} ds_i}{b-v})\}.$$

## 4.3 Model 2: Centralized System Model (CSM)

Different from DSM, the supply chain of CSM adds a DC that takes orders from all retailers. The DC location is a decision variable that determines the overall profit. Before a selling season starts, the supplier needs to decide the total order quantity $Q_0$ as well as the optimal DC location to maximize the expected profit of the whole system. When the season begins, each retailer will place an order based on its real demand. The DC ensures that each retailer will receive its full order, indicating a potential salvage or shortage. The total realized demand at DC is $x_0 = \sum_{i=1}^{N} x_i$. Assume demands at retailers are independent, therefore $x_0$ also follows the normal distribution with a mean $\mu_0 = \sum_{i=1}^{N} \mu_i$ and a standard deviation $\sigma_0 = \sqrt{\sum_{i=1}^{N} \sigma_i^2}$.

We formulate the overall expected profit, $\Pi^C(Q_0, \mathcal{X})$, as follows:

$$\Pi^C(Q_0, \mathcal{X}) = s \int_0^\infty x_0 f(x_0) dx_0 - b \int_{Q_0}^\infty (x_0 - Q_0) f(x_0) dx_0$$
$$+ v \int_0^{Q_0} (Q_0 - x_0) f(x_0) dx_0 - cQ_0 - \sum_{i \in \mathcal{I}} K_i n$$
$$- \sum_{i \in \mathcal{I} \cup \{0\}} TI_i(x_i, d_i) - T0(Q_0, d_0)$$

subject to
$$Q_0 \geq \sum_{i \in \mathcal{I}} Q_i,$$
$$F_i(Q_i) \geq \gamma, \quad \forall i \in \mathcal{I},$$
$$d_0(x, y) \geq 0,$$
$$d_i(x, y) \geq 0, \quad \forall i \in \mathcal{I}.$$

In this formulation, the expected profit is a function of total order quantity $Q_0$ and the DC location $(X, y)$, which in return impacts the distances $d_i$, $i \in \mathcal{I}$ and $d_0$. The first part of the objective function captures the newsvendor related revenue and cost, including total revenue, shortage cost and salvage revenue, production cost of supplier, and fixed replenishment cost at each retailer. Again, we assume $K_i = 0$. The second part is the transportation cost from supplier to DC and then DC to retailers. In CSM we assume that there is no transshipment among retailers. Since the shipments happened after receiving accurate demand information from retailers, the quantity shipped to each retailer is equal to the actual demand $x_i$. The distances between facilities can be calculated using squared Euclidean, Euclidean, or Rectangular distances. Similar to DSM, CSM also requires that the quantity planned to distribute to each retailer $i$, i.e., $Q_i$ needs to satisfy its minimal service level $F_i^{-1}(\gamma)$.

## 4.4 Solutions of CSM

In this section, we discuss the solution approaches of CSM under three types of transportation costs. Compared with DSM, there are gains from risk-pooling and centralization of the inventory. But the transportation cost could consume the risk-pooling benefit. We will compare DSM and CSM in next section.

Same as DSM, we consider three different transportation costs functions in CSM.

**Case 1:** $T0(Q_0, d_0) = p_0^q + r_0^q Q_0 \$, \$ TI_i(Q_i, d_i) = p_i^q + r_i^q Q_i$

For case 1, the DC location has no impact on the expected profit as it is not included in the model. The unit transportation cost $r_0$, however, will affect the optimal order quantity $Q_0$, and therefore, influence the total profit. We assume that the unit transportation costs $r_i^q$ from DC to retailer $i$ are the same for all retailers, i.e., $r_i^q = r^q$. Since the final quantity of products delivered to retailers equals to the actual demands, the transportation cost from the DC to retailers is unrelated to the optimal order quantity $Q_0$. Then we obtain the optimal central order quantity $Q_0^*$:

$$Q_0^* = \max\{\sum_i F_i^{-1}(\gamma), F_0^{-1}(\frac{b-c-r_0^q}{b-v})\}$$

**Case 2:** $T0(Q_0, d_0) = p_0^d + r_0^d d_0 \$, \$ TI_i(Q_i, d_i) = p_i^d + r_i^d d_i$

When the transportation cost depends on the travel distance, we can separate the objective function to two independent parts. The first part is the newsvendor problem of DC which only contains decision variable $Q_0$, and the second part is the location problem to determine the optimal DC location. We can easily find the optimal order quantity at DC is:

$$Q_0^* = \max\{\sum_i F_i^{-1}(\gamma), F_0^{-1}(\frac{c_u}{c_u + c_o}) = F_0^{-1}(\frac{b-c}{b-v})\}$$

For the location problem we want to minimize the total weighted travel distance of the entire supply chain network:

$$\text{minimize} \quad \sum_{i \in \mathcal{I}} TI_i(Q_i, d_i) + T0(Q_0, d_0) = \sum_{i \in \mathcal{I} \cup \{0\}} (p_i^d + r_i^d d_i)$$

in which:

$p_i^d$ is the bundling cost that is associated with one shipment,

$r_i^d$ is the unit transportation cost,

$X = (x, y)$ is the location of DC,

$L_i = (a_i, b_i)$ ) is the location of retailer $i$,

$L_0 = (a_0, b_0)$ is the location of supplier,

$d_0$ is the distance between supplier and DC,

$d_i$ is the distance between DC and retailer $i$, $\forall i \in \mathcal{I}$,

$d_i = \quad [(x - a_i)^2 + (y - b_i)^2 + \epsilon]^{1/2}$.

Clearly, the center-of-gravity is the optimal solution for this subproblem. We have the optimal DC location:

$$x = \frac{\sum_{i \in \mathcal{I} \cup \{0\}} r_i^d a_i}{\sum_{i \in \mathcal{I} \cup \{0\}} r_i^d}, \quad y = \frac{\sum_{i \in \mathcal{I} \cup \{0\}} r_i^d b_i}{\sum_{i \in \mathcal{I} \cup \{0\}} r_i^d}.$$

**Case 3:** $T0(Q_0, d_0) = p_0^{qd} + r_0^{qd} Q_0 d_0$, $TI_i(Q_i, d_i) = p_i^{qd} + r_i^{qd} Q_i d_i$

In this transportation mode, the total transportation cost is related to both $Q_0$ and $d_0$, which can be represented as:

$$\sum_{i \in \mathcal{I}} TS_i(x_i, d_i) + T0(Q_0, d_0) = \sum_{i \in \mathcal{I}} [p_i^{qd} + r_i^{qd} x_i * [(x - a_i)^2 + (y - b_i)^2 + \epsilon]^{1/2}]$$
$$+ [p_0^{qd} + r_0^{qd} Q_0 * [(x - a_0)^2 + (y - b_0)^2 + \epsilon]^{1/2}]$$

The separation applied to case 2 is no longer feasible for case 3 due to the nonlinearity of this new form of transportation cost. To this end, we need to examine the concavity of the objective function to determine a feasible solution approach. Let $H(\Pi^C)$ be the Hessian Matrix of $\Pi^C$.

**Theorem 1.** *With the demand $x_0$ following normal distribution, $\Pi^C$ is NOT concave on all decision variables, because $H(\Pi^C)$ is NOT negative definite.*

Proof. $H(\Pi^C)$ is negative definite if and only if the scalar $z^T M z$ is negative for every non-zero column vector $z$ of real numbers. Therefore, to show that $H(\Pi^C)$ is NOT negative definite, we only need to find one particular instance of $z$ such that $z^T H z$ is positive. To simplify the calculation, we consider a network with a supplier, a DC, and a retailer, i.e., $n=1$. Thus $\Pi^C$ becomes a function of three variables including $Q_0$, $x$, and $y$, and $H(\Pi^C)$ is a $3\times 3$ matrix. In addition, let $p = p_0 = p_i$, $r = r_0 = r_i$, we write $\Pi^{C1}$ as

$$\Pi^{C1} = s\int_0^\infty \frac{x_0}{\sigma\sqrt{2\pi}} e^{-\frac{1}{2\mu^2}(x_0-\mu)^2} dx_0$$

$$+ v\int_0^{Q_0} \frac{(Q_0 - x_0)}{\sigma\sqrt{2\pi}} e^{-\frac{1}{2\mu^2}(x_0-\mu)^2} dx_0$$

$$- b\int_{Q_0}^\infty \frac{(-Q_0 + x_0)}{\sigma\sqrt{2\pi}} e^{-\frac{1}{2\mu^2}(x_0-\mu)^2} dx_0$$

$$- Q_0 c - 2p - r(Q_0 + \mu_1)\sqrt{(x-a_0)^2 + (y-b_0)^2}$$

The corresponding Hessian Matrix is

$$H(\Pi^C) = \begin{bmatrix} H_{00} & H_{01} & H_{02} \\ H_{10} & H_{11} & H_{12} \\ H_{20} & H_{21} & H_{22} \end{bmatrix}$$

in which

$$H_{00} = \frac{(v-b)b}{\sigma\sqrt{2\pi}} e^{-\frac{1}{2\mu^2}(x_0-\mu)^2}$$

$$H_{01} = -\frac{r(x-a_0)}{\sqrt{(x-a_0)^2+(y-b_0)^2}}$$

$$H_{02} = -\frac{r(y-b_0)}{\sqrt{(x-a_0)^2+(y-b_0)^2}}$$

$$H_{10} = -\frac{r(x-a_0)}{\sqrt{(x-a_0)^2+(y-b_0)^2}}$$

$$H_{11} = -\frac{Q_0 r(-x+a_0)(x-a_0)}{\left((x-a_0)^2+(y-b_0)^2\right)^{\frac{3}{2}}} - \frac{Q_0 r}{\sqrt{(x-a_0)^2+(y-b_0)^2}}$$
$$- \frac{rx_1(-x+a_1)(x-a_1)}{\left((x-a_1)^2+(y-b_1)^2\right)^{\frac{3}{2}}} - \frac{rx_1}{\sqrt{(x-a_1)^2+(y-b_1)^2}}$$

$$H_{12} = -\frac{Q_0 r(x-a_0)(-y+b_0)}{\left((x-a_0)^2+(y-b_0)^2\right)^{\frac{3}{2}}} - \frac{rx_1(x-a_1)(-y+b_1)}{\left((x-a_1)^2+(y-b_1)^2\right)^{\frac{3}{2}}}$$

$$H_{20} = -\frac{r(y-b_0)}{\sqrt{(x-a_0)^2+(y-b_0)^2}}$$

$$H_{21} = -\frac{Q_0 r(x-a_0)(-y+b_0)}{\left((x-a_0)^2+(y-b_0)^2\right)^{\frac{3}{2}}} - \frac{rx_1(x-a_1)(-y+b_1)}{\left((x-a_1)^2+(y-b_1)^2\right)^{\frac{3}{2}}}$$

$$H_{22} = -\frac{Q_0 r(-y+b_0)(y-b_0)}{\left((x-a_0)^2+(y-b_0)^2\right)^{\frac{3}{2}}} - \frac{Q_0 r}{\sqrt{(x-a_0)^2+(y-b_0)^2}}$$
$$- \frac{rx_1(-y+b_1)(y-b_1)}{\left((x-a_1)^2+(y-b_1)^2\right)^{\frac{3}{2}}} - \frac{rx_1}{\sqrt{(x-a_1)^2+(y-b_1)^2}}$$

Let $z = [z_0, z_1, z_2]^T$, we have

$$z^T Hz = -z_0 \frac{rz_1(x-a_0)+rz_2(y-b_0)}{\sqrt{(x-a_0)^2+(y-b_0)^2}} + \frac{z_0^2(v-b)b}{\sigma\sqrt{2\pi}} e^{-\frac{1}{2\mu^2}(x_0-\mu)^2}$$

$$-z_1 \frac{rz_0(x-a_0)}{\sqrt{(x-a_0)^2+(y-b_0)^2}} - z_1^2 \frac{Q_0 r(-x+a_0)(x-a_0)}{\left((x-a_0)^2+(y-b_0)^2\right)^{\frac{3}{2}}}$$

$$-z_1^2 \frac{Q_0 r}{\sqrt{(x-a_0)^2+(y-b_0)^2}} - z_1^2 \frac{rx_1(-x+a_1)(x-a_1)}{\left((x-a_1)^2+(y-b_1)^2\right)^{\frac{3}{2}}}$$

$$-z_1^2 \frac{rx_1}{\sqrt{(x-a_1)^2+(y-b_1)^2}} - z_1 z_2 \frac{Q_0 r(x-a_0)(-y+b_0)}{\left((x-a_0)^2+(y-b_0)^2\right)^{\frac{3}{2}}} n$$

$$-z_1 z_2 \frac{rx_1(x-a_1)(-y+b_1)}{\left((x-a_1)^2+(y-b_1)^2\right)^{\frac{3}{2}}} - z_2 \frac{rz_0(y-b_0)}{\sqrt{(x-a_0)^2+(y-b_0)^2}} n$$

$$-z_1 z_2 \frac{Q_0 r(x-a_0)(-y+b_0)}{\left((x-a_0)^2+(y-b_0)^2\right)^{\frac{3}{2}}} - z_1 z_2 \frac{rx_1(x-a_1)(-y+b_1)}{\left((x-a_1)^2+(y-b_1)^2\right)^{\frac{3}{2}}} n$$

$$-z_2^2 \frac{Q_0 r(-y+b_0)(y-b_0)}{\left((x-a_0)^2+(y-b_0)^2\right)^{\frac{3}{2}}} - z_2^2 \frac{Q_0 r}{\sqrt{(x-a_0)^2+(y-b_0)^2}}$$

$$-z_2^2 \frac{rx_1(-y+b_1)(y-b_1)}{\left((x-a_1)^2+(y-b_1)^2\right)^{\frac{3}{2}}} - z_2^2 \frac{rx_1}{\sqrt{(x-a_1)^2+(y-b_1)^2}}$$

Now we feed some real numbers into the expression. Let $s=200$, $c=50$, $v=20$, $b=25$, $p=100$, $r=0.05$, $a_0=100$, $b_0=100$, $a_1=500$, $b_1=500$, $x=300$, $y=300$, $Q_0=1000$, $\mu=100$, $\sigma=10$, $z=[-100,1,1]^T$. We have

$$z^T Hz = -\frac{2500\sqrt{2}}{\sqrt{\pi}e^{4050}} + 10.0\sqrt{2} \approx 0 + 10.0\sqrt{2} > 0$$

Therefore, $H(\Pi^C)$ is not negative definite, and $\Pi^C$ is not concave. □

### 4.4.1 Solution Approach: Q-search

Because the profit is not concave on all decision variables, it is not viable to solve the problem with the augmented Lagrange multiplier method. To this end, we develop a custom solution algorithm named Q-search that can identify a **good** solution that is locally optimal. We find that when $Q_0$ is given, the problem is reduced to a simplified non-linear problem with the DC location $(x, y)$ as decision variables, which can be solved by the sequential least squares programming (SLSQP) algorithm. The proposed Q-search algorithm explores a wide range of possible $Q_0$ s, evaluates the reduced non-linear problem via the SLSQP optimizer, stores the solution of each evaluation, and eventually outputs the one with highest profit as the solution we accept. In particular, the algorithm is described as follows:

**Step 1**: Determine the lower limit of $Q_0$, $Q_{lb} = \sum_{i \in \mathcal{I}} F_i^{-1}(\gamma)$, based on the service level constraint, and use it as a starting point of the search.

**Step 2**: Given the current $Q_0$, solve the simplified non-linear problem to obtain a profit, store tuple $(profit, Q_0, x, y)$ in the result set.

**Step 3**: Increment $Q_0$ by a pre-set step size, if $Q_0$ reaches $2Q_{lb}$, stop the search. Otherwise go to step 2.

**Step 4**: Return the entry with highest profit in the result set.

## 4.5 Choosing a Retailer as DC

Rather than building a new DC, an alternative is to set the DC at the closet retailer to the optimal DC location. Both options will incur additional cost, but in practice the cost of the retailer-as-DC option should be less expensive. We will explore this strategy via experiment in Section 6.

# 5 A Comparison of DSM and CSM

In this section, we provide a comparison of DSM and CSM under the first two types of transportation cost.

## 5.1 Case 1: quantity-dependent transportation cost

In Section 4.2 and Section 4.4, we have shown the optimal solutions for both DSM and CSM:

$$Q_i^* = \max\{F_i^{-1}(\gamma), F_i^{-1}(\frac{b-c-rs_i^q}{b-v})\} \quad (DSM)$$

$$Q_0^* = \max\{\sum_i F_i^{-1}(\gamma), F_0^{-1}(\frac{b-c-r_0^q}{b-v})\} \quad (CSM)$$

In DSM, the total optimal order quantity is $\sum_{i \in \mathcal{I}} Q_i$. Let $\beta_i = \frac{b-c-rs_i^q}{b-v}, \beta_0 = \frac{b-c-r_0^q}{b-v}$, then we can write the second expression of the max function in both solutions as $F_i^{-1}(\beta_i)$ and $F_0^{-1}(\beta_0)$. We first focus on the comparison when the optimal solutions are $F_i^{-1}(\beta_i)$ and $F_0^{-1}(\beta_0)$. This applies $\gamma \geq \beta_i$, and $\gamma \geq \beta_0$.

The normal distribution assumption is widely used in inventory theories when the demand comes from many different independent or weakly dependent retailers. In this paper, we assume that the demand of each retailer follows normal distribution $N(\mu_i, \sigma_i)$, then we can write the demand of retailer $i$ as $x_i = \mu_i + Z\sigma_i$ where $Z$ is a standard normal random variable, and $\Phi(z) = Pr(Z < z)$ is the CDF of the standard normal random variable. Let $z_{\beta_i} = \Phi^{-1}(\beta_i)$ and $z_{\beta_0} = \Phi^{-1}(\beta_0)$, we can get

$$Q_i^* = \mu_i + z_{\beta_i}\sigma_i, Q_0^* = \mu_0 + z_{\beta_0}\sigma_0.$$

To simplify the objective function and let readers go through the results easily, we assume the unit transportation costs are the same for all retailers, $rs_i^q = rs^q, r_i^q = r^q \forall i \in \mathcal{I}$. With the optimal order quantity, the profit gained from DSM is:

$$\Pi^D = (b-c-rs^q)\sum \mu_i - \sum G(Q_i^*)$$

where

$$\sum_{i\in\mathcal{I}} G(Q_i^*) = \sum_{i\in\mathcal{I}}[(c+rs^q-v)E[Q_i^*-x_i]^+ + (b-c-rs^q)E[x_i-Q_i^*]^+]$$
$$= (c+rs^q-v)\sum_{i\in\mathcal{I}}Q_i^* - (c+rs^q-v)\sum_{i\in\mathcal{I}}\mu_i + (b-v)E\sum_{i\in\mathcal{I}}[x_i-Q_i^*]^+$$
$$= (c+rs^q-v)\sum_{i\in\mathcal{I}}Q_i^* - (c+rs^q-v)\sum_{i\in\mathcal{I}}\mu_i + (b-v)E\sum_{i\in\mathcal{I}}\sigma_i R(z_{\beta_i})$$

We define $R(u) = Pr(z>u) = \int_u^\infty (x-u)\frac{1}{\sqrt{2\pi}}e^{-x^2/2}dx$, as the right-hand unit normal linear-loss integral.

Similarly, we can show the profit gained with CSM is:

$$\Pi^C = (b-c-r_0^q)\mu_0 - G(Q_0^*)$$

where

$$G(Q_0^*) = (c+r_0^q-v)E[Q_0^*-x_0]^+ + (b-c-r_0^q)E[x_0-Q_0^*]^+$$
$$= (c+r_0^q-v)Q_0^* - (c+r_0^q-v)\mu_0 + (b-v)E[x_0-Q_0^*]^+$$
$$= (c+r_0^q-v)Q_0^* - (c+r_0^q-v)\mu_0 + (b-v)\sigma_0 R(z_{\beta_0}).$$

For simplicity, assume that $\sigma_i$ is constant over all retailers, i.e., $\sigma_i = \sigma$, and $\sum_{i\in\mathcal{I}} Q_i^* = Q_0^*$.

Then the difference of profit between CSM and DSM is:

$$\Pi^C - \Pi^D = (s-c-r_0^q-r^q)\mu_0 - G(Q_0^*) - (s-c-rs^q)\sum_{i\in\mathcal{I}}\mu_i + \sum_{i\in\mathcal{I}}G(Q_i^*)$$
$$= (s-c-r_0^q-r^q)\mu_0 - (s-c-rs^q)\mu_0 - G(Q_0^*) + \sum_{i\in\mathcal{I}}G(Q_i^*)$$
$$= (rs-r_0^q-r^q)\mu_0 + (c+rs^q-v)\sum_{i\in\mathcal{I}}Q_i^* - (c+rs^q-v)\mu_0 + (b-v)\sum_{i\in\mathcal{I}}\sigma_i R(z_{\beta_i})$$
$$\quad - (c+r_0^q-v)Q_0^* + (c+r_0^q-v)\mu_0 - (b-v)\sigma_0 R(z_{\beta_0})$$
$$= -r_0^q\mu_0 + (c-v)(\sum_{i\in\mathcal{I}}Q_i - Q_0) + rs^q\sum_{i\in\mathcal{I}}Q_i - r_0^q Q_0$$
$$\quad + (b-v)[\sum_{i\in\mathcal{I}}\sigma_i R(z_{\beta_i}) - \sigma_0 R(z_{\beta_0})]$$
$$= (rs-r_0^q-r^q)\mu_0 + (c-v)(\sum_{i\in\mathcal{I}}\sigma_i z_{\beta_i} - \sigma_0 z_{\beta_0}) + rs^q\sum_{i\in\mathcal{I}}\sigma_i z_{\beta_i} - r_0^q \sigma_0 z_{\beta_0}$$
$$\quad + (b-v)\sqrt{n}\sigma[\sqrt{n}R(z_{\beta_i}) - R(z_{\beta_0})]$$

Now, we can clearly see the impact of unit transportation cost on the profit. Based on the relationship between $rs^q$ and $r_0^q + r^q$, the benefit of using CSM may vary:

**Lemma 1.** *If $\delta = rs^q - r_0^q > 0$ and $\sqrt{n} z_{\beta_i} \geq z_{\beta_0}$, then profit of CSM is always greater than the profit of DSM. The difference of $\Pi^C - \Pi^D$ is positive related to $\delta, n$ and $\sigma$.*

*Proof.* If $rs^q > r_0^q + r^q$: Let $\delta = rs^q - r_0^q - r^q > 0$, and recall that $\beta_i = \dfrac{b-c-rs^q}{b-v}$ and $\beta_0 = \dfrac{b-c-r_0^q}{b-v}$, we have

$$\beta_i < \beta_0$$
$$\Rightarrow z_{\beta_i} < z_{\beta_0},$$
$$\Rightarrow R(z_{\beta_i}) > R(z_{\beta_0}),$$
$$\Rightarrow (b-v)\sqrt{n}\sigma[\sqrt{n} R(z_{\beta_i}) - R(z_{\beta_0})].$$

If $\sqrt{n} z_{\beta_i} \geq z_{\beta_0}$, we can have

$$\sqrt{n}\sigma_i(\sqrt{n} z_{\beta_i} - z_{\beta_0}) \geq 0,$$
$$\Rightarrow \sum_{i \in \mathcal{I}} \sigma_i z_{\beta_i} - \sigma_0 z_{\beta_0} \geq 0.$$

Therefore, when $\delta > 0$, and $\Pi^C - \Pi^D > 0$. The difference is related to the value of $\delta, n$ and $\sigma$. It implies that when the CSM is more profitable by 1) saving more from central transportation, 2) having more retailers, and 3) increasing demand variation at retailers. For a special case, when $\delta = 0$, meaning there is no difference on the unit transportation cost in two models, CSM is still a better choice because of the risk-pooling benefit. □

This case is similar to the previous work by Eppen and others (Cherikh, 2000; Eppen, 1979; Lin et al., 2001). However, when $\delta < 0$, CSM make more profit only if $(s-v)\sqrt{n}\sigma[\sqrt{n} R(z_{\beta_i}) - R(z_{\beta_0})] > (r_0^q + r^q - rs^q)\mu_0^*$, which can only be true when the number of retailers is very large.

When $\gamma > \beta_i$ and $\gamma > \beta_0$, we can find the optimal order quantity of DSM and CSM are:

$$Q_i^* = \mu_i + z_\gamma \sigma_i, Q_0^* = \sum Q_i.$$

Therefore, the difference of profit between CSM and DSM is

$$\Pi^C - \Pi^D = (s - c - r_0^q - r^q)\mu_0 - G(Q_0^*) - (s - c - rs^q)\sum_{i \in \mathcal{I}} \mu_i + \sum_{i \in \mathcal{I}} G(Q_i^*)$$

$$= -r^q \mu_0 + (rs^q - r_0^q)Q_0 \sigma + (b - v)\sqrt{n}\sigma R(z_\gamma)[\sqrt{n} - 1]$$

When $rs^q > r_0^q + r^q$, clearly this difference is positive. It means CSM outperforms DSM.

## 5.2 Case 2: distance-dependent transportation cost

With distance-related transportation cost, we have shown the optimal solutions for DSM and CSM in the previous section.

$$Q_i^* = \max\{F_i^{-1}(\gamma), F_i^{-1}(\frac{b-c}{b-v})\} \quad \forall i \in \mathcal{I} \quad (DSM)$$

$$Q_0^* = \max\{\sum_i F_i^{-1}(\gamma), F_0^{-1}(\frac{c_u}{c_u + c_o}) = F_0^{-1}(\frac{b-c}{b-v})\} \quad (CSM)$$

Letting $\frac{b-c}{b-v} = \beta, \sum_i \mu_i = \mu_0$, and $\sigma_i = \sigma, \forall i \in \mathcal{I}$, the difference of revenue and inventory costs between CSM and DSM is

$$CSM - DSM = (s - c)\mu_0 - G(Q_0^*) - (s - c)\sum_{i \in \mathcal{I}} \mu_i + \sum_{i \in \mathcal{I}} G(Q_i^*)$$

$$= \sum_{i \in \mathcal{I}} G(Q_i^*) - G(Q_0^*)$$

$$= (c - v)(\sum_{i \in \mathcal{I}} Q_i - Q_0) + (b - v)[\sum_{i \in \mathcal{I}} \sigma_i R(z_i) - \sigma_0 R(z_0)].$$

If $Q_i^* = \sum_i F_i^{-1}(\gamma), Q_0^* = \sum_i F_i^{-1}(\gamma)$, then $\sum_{i \in \mathcal{I}} Q_i = Q_0$. Because $\sigma_i$ is constant for all retailers, we have

$$\Pi^C - \Pi^D = (b-v)[\sum_{i\in\mathcal{I}}\sigma_i R(z_i) - \sigma_0 R(z_0)]$$
$$= (b-v)(n\sigma - \sqrt{n}\sigma)R(z)$$
$$= (b-v)\sqrt{n}\sigma(\sqrt{n}-1)R(z) > 0$$

This shows that CSM benefits from risk pooling, and the profit is positively related to the variations of retailer demand. CSM also works better if the number of retailers increases. However the extra earnings from centralization may reduce if we consider the transportation costs. In case 2, the transportation cost depends on the total distance traveled. Using a central DC will incur longer distance compared with a direct shipping from the supplier.

The total transportation cost in CSM is

$$\sum_{i\in\mathcal{I}}(r_0^d d_0 + r_i^d d_i)$$

and the total transportation cost in DSM is

$$\sum_{i\in\mathcal{I}}(rs_i^d ds_i)$$

Since we want to maximize the total profit, the transportation part needs to be minimized. If we assume that $r_0^d = r_i^d = rs_i^d = r^d$, clearly DSM takes shorter distances than CSM, regardless of the DC location. The difference between CSM and DSM on transportation is

$$Trans(CSM - DSM) = \sum r^d(d_0 + d_i - ds_i) > 0.$$

It shows that in case 2, the performance of CSM may be worse than DSM. It depends on the difference of risk-pooling benefits and extra transportation cost.

## 5.3 Case 3: quantity-distance-dependent transportation cost

We will examine the performance of DSM and CSM under a quantity-distance-dependent transportation cost through computational experiments in next section.

# 6 Computational Experiments

In this section, we only focus on case 3 for the DSM and CSM, as we have compared the two models for case 1 and case 2 with numeric analysis in Section 5. We conduct all computational experiments on a workstation with an Intel Core (TM) i7-4770 CPU (3.40GHz), a 16GB RAM, and a 64-bit Windows 10 operation system. The employed solver is the optimization package of Scipy, an open-source Python library for scientific computation. In particular, we utilize the built-in SLSQP optimizer in Scipy as a building block of our solution approach. We implement both the data generation and the solution approach in Python.

Table 1: Data Specification

| Parameter | Notation | Setting |
|---|---|---|
| # Retailers | $n$ | [10, 100] by 10 |
| Selling Price | $s$ | $200 |
| Unit Cost | $c$ | $50 |
| Unit Salvage Value | $v$ | $20 |
| Map Size | map_sz | 1000 × 1000 |
| Bundling Cost | $(p_0, p_i)$ | ($200, $100) |
| Unit Trans. Cost | $(r_0^{qd}, r_i^{qd})$ | ($0.3, 0.5$) |
| Demand Mean | $\mu_i$ | U[100, 200] |
| Demand Stdev | $\sigma_i$ | U[10, 20] |
| Pre-set Service Level | $\gamma$ | 0.3 |

## 6.1 Data Setting

We present the data setting of the experiment in Table 1, which includes ten instances that differ in network size measured by the number of retailers. The tested supply chain networks include 10, 20, up to 100 retailers, which can cover a wide range of real world situations. The demand of each retailer follows a normal distribution, with a mean in [100, 200] and a standard deviation in [10, 20]. We consider a square map with an area of 1000 × 1000 square miles; all location coordinates are randomly generated within the map. A fixed seed for the random number generator is used throughout the experiment for the purpose of consistency. In addition, we use $\gamma=0.3$ for the tested ten instances.

## 6.2 DSM and CSM: a Comparison for Case 3

In this section, we conduct a comparison study of case 3 for DSM and CSM. Letting $D_i$ denote the real demand retailer $i$ observes after the selling season starts, we define four metrics to measure the demand fulfillment rate:

Table 2: DSM Vs. CSM: Optimal Expected Profits

| n | DSM | | CSM | | $\Delta$ ($) |
|---|---|---|---|---|---|
| | $P_{DSM}$ ($) | $M_1$ | $P_{CSM}$ ($) | $M_3$ | |
| 10 | 168879 | 94.6% | 170164 | 98.3% | 1285 |
| 20 | 354684 | 94.9% | 357101 | 99.3% | 2417 |
| 30 | 532537 | 94.8% | 536232 | 99.4% | 3695 |
| 40 | 688633 | 94.7% | 698628 | 99.1% | 9994 |
| 50 | 915006 | 94.6% | 924055 | 99.5% | 9049 |
| 60 | 1155430 | 94.6% | 1167504 | 99.7% | 12075 |
| 70 | 1274804 | 94.6% | 1294238 | 99.4% | 19434 |
| 80 | 1437268 | 94.5% | 1455582 | 99.5% | 18314 |
| 90 | 1612475 | 94.4% | 1629188 | 99.6% | 16713 |
| 100 | 1639821 | 94.3% | 1689876 | 99.2% | 50056 |

$$DSM: M_1 = \frac{1}{n}\sum_{i \in \mathcal{I}} \frac{Q_i}{\mu_i} \quad M_2 = \frac{1}{n}\sum_{i \in \mathcal{I}} \frac{Q_i}{D_i}$$

$$CSM: M_3 = \frac{Q_0}{\sum_{i \in \mathcal{I}} \mu_i} \quad M_4 = \frac{Q_0}{\sum_{i \in \mathcal{I}} D_i}$$

in which $M_1$ and $M_3$ reflect the ratio of order size to the expected demand, so they are referred to as expected fulfillment, while $M_2$ and $M_4$ measure the ratio of order size to the real demand, so we call them actual demand fulfillment.

Table 2 presents the optimal profits, the service level indicator $M_1$, the running time for the CSM as well as the profit difference for both systems. In the table header, $P_{DSM}$ and $P_{CSM}$ refer to the optimal expected profits for DSM and CSM, respectively; also, we have $\Delta = P_{CSM} - P_{DSM}$, as shown in the last column. We find that for all ten instances, CSM outperforms DSM in terms of both profit and service level. Also, the profit gap $\Delta$ measures the additional profit the CSM earns compared to DSM. As setting a DC at a location incurs a cost, which can be taken into account at this point: if $\Delta$ is positive, and the amount is greater than the cost of adding a DC, then the CSM is a preferred choice. In this table, when the system size becomes larger, $\Delta$ is also larger, meaning that the CSM is more suitable for large supply chain networks. Meanwhile, the service level indicators, i.e., $M_1$ and $M_3$, of CSM are higher than those of DSM, meaning that the CSM system can deal with a large market.

Table 3 displays the profits calculated with randomly generated demands. As both the DSM CSM are stochastic models that optimize expected profits, it is desirable to validate the models with random demand samples, which can simulate the demand observed at the beginning of the selling season. For our problem, we generate a random demand sample for each retailer using its density function, feed them into the profit function to replace $Q_i/\mu_i$, and obtain the actual profit. We find that both CSM and DSM models perform as expected with these simulated demands. The results are in line with those in Table 2.

Table 3: DSM Vs. CSM: Profits with Random Demand Samples

| | DSM | | CSM | | |
|---|---|---|---|---|---|
| n | $P_{DSM}$ ($) | $M_2$ | $P_{CSM}$ ($) | $M_4$ | $\Delta$ ($) |
| 10 | 172419 | 100.3% | 175578 | 102.9% | 3159 |
| 20 | 361815 | 100.7% | 366601 | 104.1% | 4786 |
| 30 | 542087 | 99.8% | 546837 | 103.3% | 4749 |
| 40 | 700733 | 99.4% | 711599 | 102.8% | 10866 |
| 50 | 931974 | 100.1% | 942193 | 103.8% | 10219 |
| 60 | 1174275 | 99.7% | 1183432 | 103.6% | 9157 |
| 70 | 1297803 | 99.5% | 1314126 | 103.2% | 16323 |
| 80 | 1459912 | 98.8% | 1477041 | 102.8% | 17128 |
| 90 | 1635625 | 98.3% | 1650692 | 102.6% | 15067 |
| 100 | 1669819 | 98.9% | 1719961 | 102.7% | 50142 |

## 6.3 Retailer-as-DC: a Potentially Better Option

We discuss the option of choosing a retailer closest to the optimal DC location to serve as a temporary DC, which is more economic to build and maintain. We evaluate this retailer - as-DC option and compare it with the optimal solution. We study one instance (n=10) and plot the result in Figure 2, in which the horizontal axis shows the location of the optimal DC and locations of ten retailers, and the vertical axis refers to overall profit. In this case, as an alternative, we can build the DC at retailer 3, the one closest to the optimal DC, and live with a profit loss of $4,873. However, building a DC at the optimal location costs more than building it at a retailer. In practice, if this cost difference is over $4,873, retailer-as-DC wins. We expand this experiment to all ten instances. Table 4 shows the optimal DC location and the location of the closest retailer, and Table 5 displays a comparison of the two options interms of expected and actual profits. Note that $P_{CSM'}$ refers to the profit obtained based on the location of the chosen retailer.

The first finding is that the system scale has no clear impact on the profit gap, which is mainly determined by the distance between the optimal DC and its closest retailer. Interestingly, for one instance (n=70), we obtain a negative profit gap, meaning that retailer-as-DC makes more actual profit with the given demand samples. This is because the solver uses expected demands to compute an optimal expected profit and a DC location, while the actual profit relies on simulated demands. Therefore, our models can provide valuable guidance but cannot guarantee an optimal decision provided with random demand samples, because of the nature of the problem. Overall, we highly recommend retailer-as-DC if the chosen retailer is close to the optimal DC location.

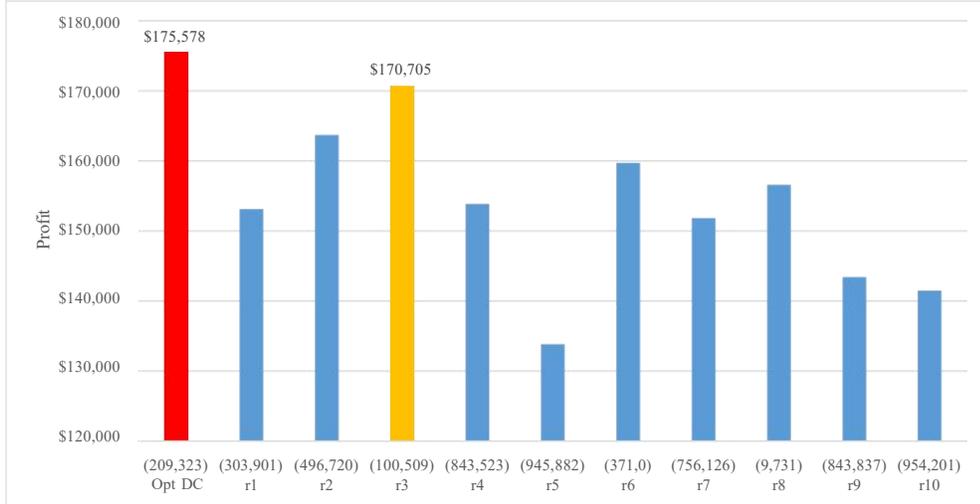

Figure 2: Optimal DC Vs. Retailer-as-DC on Instance n=10

## 6.4 Parameter Sensitivity Analysis

For sensitivity analysis, we select a mid-size instance with 40 retailers from the ten instances that are previously generated. For each experiment we only vary one parameter and keep the rest parameters unchanged. Table 6 displays three parameters including the pre-set service level, the map size, and the unit transportation cost, as well as their values in the experiment.

Table 4: CSM Case 3: Optimal DC Vs. Retailer-as-DC

| n | $Q_0^*$ | $DC^*$ | Chosen Retailer | Dist. (mi) |
|---|---|---|---|---|
| 10 | 1467 | (209.6, 323.0) | (100, 509) | 215.9 |
| 20 | 2935 | (121.0, 144.1) | (90, 62) | 87.8 |
| 30 | 4421 | (496.7, 915.7) | (451, 905) | 46.9 |
| 40 | 5898 | (250.6, 205.0) | (206, 197) | 45.3 |
| 50 | 7435 | (847.2, 359.5) | (854, 345) | 16.0 |
| 60 | 9019 | (269.0, 360.0) | (308, 398) | 54.5 |
| 70 | 10576 | (633.8, 177.0) | (650, 144) | 36.8 |
| 80 | 11990 | (186.7, 702.3) | (200, 653) | 51.1 |
| 90 | 13272 | (735.0, 773.5) | (725, 792) | 21.0 |
| 100 | 14493 | (237.3, 228.5) | (251, 233) | 14.4 |

Table 5: CSM Profit Comparison: Optimal DC Vs. Retailer-as-DC

| n | Expected Profit | | | Actual Profit | | |
|---|---|---|---|---|---|---|
| | $P_{CSM}(\$)$ | $P_{CSM'}(\$)$ | $\Delta(\$)$ | $P_{CSM}(\$)$ | $P_{CSM'}(\$)$ | $\Delta(\$)$ |
| 10  | 170164  | 164893  | 5271 | 175578  | 170705  | 4872 |
| 20  | 357101  | 349242  | 7859 | 366601  | 359161  | 7441 |
| 30  | 536232  | 533343  | 2889 | 546837  | 544110  | 2727 |
| 40  | 698628  | 697833  | 794  | 711599  | 711230  | 369  |
| 50  | 924055  | 923623  | 432  | 942193  | 941918  | 275  |
| 60  | 1167504 | 1159312 | 8193 | 1183432 | 1174974 | 8458 |
| 70  | 1294238 | 1293777 | 461  | 1314126 | 1314151 | -26  |
| 80  | 1455582 | 1453261 | 2321 | 1477041 | 1474350 | 2690 |
| 90  | 1629188 | 1627938 | 1250 | 1650692 | 1649627 | 1066 |
| 100 | 1689876 | 1689753 | 123  | 1719961 | 1719627 | 334  |

Table 6: Sensitivity Analysis

| Parameter | Notation | Setting |
|---|---|---|
| Service Level Limit | $\gamma$ | [0.1, 0.9] by 0.1 |
| Map Size | map_sz | [200, 400, 800, 1000] |
| Unit Trans. Cost | $(r_0^{qd}, r_i^{qd})$ | [0.03,0.05], [0.05,0.05] |

### 6.4.1 Service Level

Parameter $\gamma$ sets a lower limit for the system service level, i.e., $F_i(Q_i) \geq \gamma$; in other words, $Q_i$ takes a value that has a probability of $\gamma$ to satisfy a random demand with a CDF of $F_i$. Clearly, the higher $\gamma$ is, the larger $Q_i$ has to be. To figure out the impact of $\gamma$, we vary its value from 0.1 to 0.9 by 0.1, and display the result in Figure 3, from which we see a clear rising trend of service level and a descending trend of profit, for both DSM and CSM, as $\gamma$ grows. First, A higher $\gamma$ leads to a higher order quantity, which fulfills more demands that are unchanged during this experiment. This explains the rising of the expected demand fulfillment $M_1$ and $M_3$, and the actual demand fulfillment $M_2$ and $M_4$. Also, the overall profit declines in all scenarios, implying a counter-effect of $\gamma$. The reason is that when the order size exceeds demands due to a high $\gamma$, salvage occurs, lowering the profit.

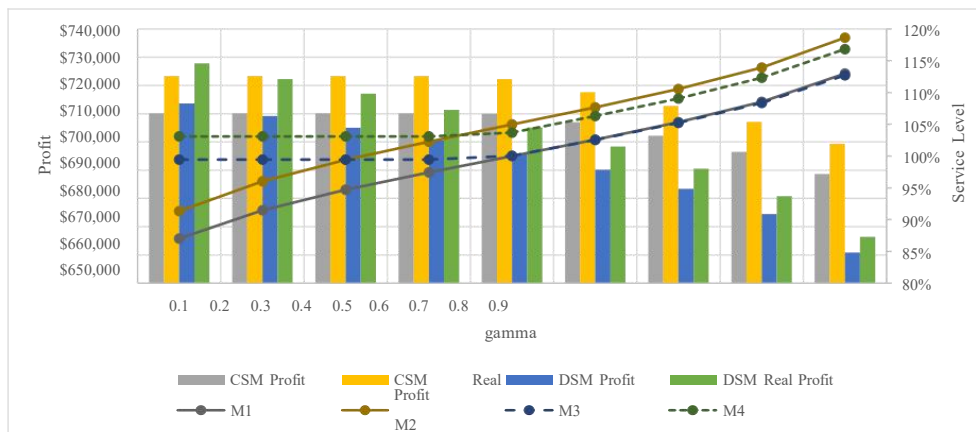

Figure 3: Impact of $\gamma$

We also find that CSM makes more profit than DSM when $\gamma$ is greater than 0.2, and the higher $\gamma$ is, the more benefit CSM creates, as the profit gap of CSM and DSM is larger. Meanwhile, we see that CSM presents a higher service level than DSM, for both expected and actual, even with a low $\gamma$. Overall, the pre-set service level does not make a positive effect for CSM, which already achieves a great performance with a $\gamma$ being 0.1; however, for DSM, a low $\gamma$ brings down both expected and actual demand fulfillment, which only boost up to 100% as $\gamma$ rises to 0.5.

### 6.4.2 Unit Transportation Cost

Our models assume that $r_0^{qd} \leq r_i^{qd}$; in other words, it is more cost-efficient to ship products to one destination (the DC) than to multiple destinations (retailers). We study the profit difference, the revenue difference, and the transportation cost difference of CSM and DSM with two modes of unit transportation cost: ($r_0^{pd}, r_i^{pd}$) = (0.03, 0.05) and ($r_0^{pd}, r_i^{pd}$) = (0.05, 0.05). Table 7 displays these differences under both expected and actual demand. With the first mode, when $r_0^{qd} \leq r_i^{qd}$, CSM incurs a higher expected transportation cost (see column three in top table) than DSM for seven out of ten instances; and this cost becomes higher when $r_0^{pd}$ is as high as $r_i^{pd}$, which can be seen from column three of the bottom table. However, the high revenue of CSM well compensates its high expense of transportation, leading to a higher profit compared to DSM, as shown in the last two columns. In addition, we should notice that CSM is less profitable with a high $r_0^{qd}$, which is illustrated by narrowed profit gaps in the bottom table.

### 6.4.3 Map Size

Intuitively, map size has a negative impact on transportation cost, since a larger map places the same number of facilities in a more scattered way, leading to increased distances among facilities. In this experiment, we raise the map from an area of 200 × 200 to 1000 × 1000. The result shown in Figure 4 verifies this intuition. Another straightforward observation is that map size is unrelated to the service level.

Table 7: Difference of CSM and DSM

| | ($r_0^{pd}, r_i^{pd}$)=(0.03, 0.05) | | | | | |
|---|---|---|---|---|---|---|
| n | $\Delta_{rev}$ | $\Delta_{trans}$ | $\Delta_{real\_rev}$ | $\Delta_{real\_trans}$ | $\Delta_{profit}$ | $\Delta_{real\_profit}$ |
| 10 | 2061 | 776 | 1641 | -1518 | 1285 | 3159 |
| 20 | 4391 | 1974 | 1861 | -2925 | 2417 | 4786 |
| 30 | 7055 | 3360 | 3143 | -1606 | 3695 | 4749 |
| 40 | 9879 | -115 | 5414 | -5452 | 9994 | 10866 |
| 50 | 12958 | 3909 | 5477 | -4742 | 9049 | 10219 |
| 60 | 15786 | 3712 | 6981 | -2176 | 12075 | 9157 |
| 70 | 18666 | -768 | 8108 | -8214 | 19434 | 16323 |
| 80 | 21542 | 3228 | 11234 | -5895 | 18314 | 17128 |
| 90 | 24542 | 7829 | 13639 | -1428 | 16713 | 15067 |
| 100 | 27153 | -22903 | 14262 | -35880 | 50056 | 50142 |
| | ($r_0^{pd}, r_i^{pd}$)=(0.05, 0.05) | | | | | |
| n | $\Delta_{rev}$ | $\Delta_{trans}$ | $\Delta_{real\_rev}$ | $\Delta_{real\_trans}$ | $\Delta_{profit}$ | $\Delta_{real\_profit}$ |
| 10 | 2096 | 1996 | 1221 | -682 | 101 | 1903 |
| 20 | 4394 | 2202 | 1711 | -2944 | 2192 | 4655 |
| 30 | 7057 | 3670 | 2993 | -1521 | 3388 | 4514 |
| 40 | 9951 | 6570 | 4604 | -13 | 3380 | 4617 |
| 50 | 12968 | 5379 | 5177 | -3844 | 7589 | 9021 |
| 60 | 15786 | 3712 | 6981 | -2176 | 12075 | 9157 |
| 70 | 18746 | 9094 | 7118 | -503 | 9652 | 7621 |
| 80 | 21596 | 10820 | 10394 | 222 | 10776 | 10171 |
| 90 | 24560 | 10809 | 13129 | 581 | 13751 | 12548 |
| 100 | 27556 | 19080 | 11772 | 279 | 8476 | 11493 |

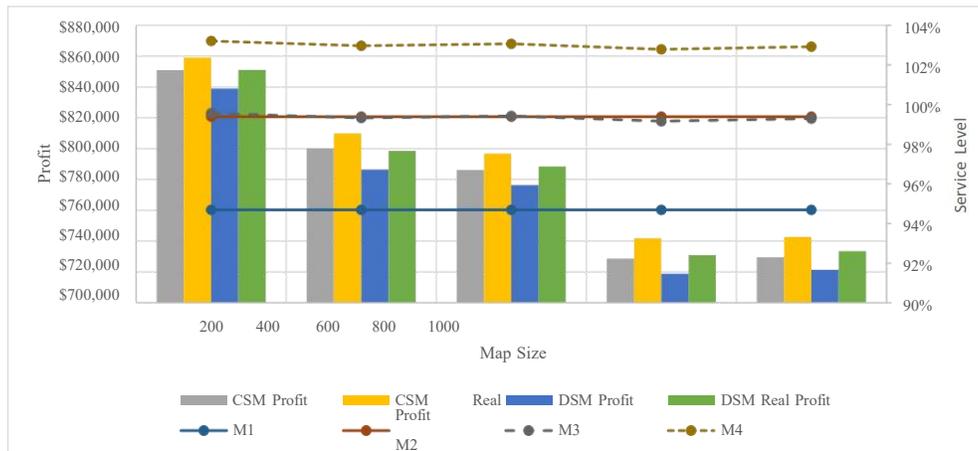

Figure 4: Impact of Map Size

# 7 Conclusion

In this paper, we consider a supply chain network where the sole supplier has the option of centralizing the ordering of a product for a number of retailers whose demand fluctuates randomly. In this network, each retailer operates in a newsvendor setting, i.e., selling a product to its customers in a single period. We discuss two models including DSM and CSM with three modes of transportation cost. For both models, the goal is to maximize the overall system profit. The DSM essentially consists of n independent newsvendor problems, while the CSM introduces a DC that takes orders from all retailers. It is thus critical to determine the location of the DC, since all orders will be firstly shipped to the DC, and then dispatched to retailers. When the transport cost is only dependent on the distance between two end points or the order size, we provide closed-form solutions for both models. However, when the transportation cost depends on both distance and order size, the CSM problem becomes a non-linear programming model, for which a closed-form optimal solution does not exist. We propose a Q-search algorithm to identify a high-quality solution which is practically acceptable. We validate our models through a series of experiments with the summarized key findings as follows. First, a comparison of DSM and CSM for case 3 shows that CSM is better for large networks in terms of both profit and service level. Also, our models are competent with simulated demands, which demonstrates their values in practice. Besides, the option of retailer-as-DC presents a great performance and is thus highly recommended if the chosen retailer is close to the optimal DC. Lastly, our sensitivity analysis shows that 1) a pre-set service level limit does not pose a positive effect on CSM; 2) the unit shipping cost ($r_0^{pd}, r_i^{pd}$) also affects the profitability of CSM, and a lower $r_0^{pd}$ helps CSM generate more profit due to more savings on shipping cost; 3) our models handles problems with various map sizes quite well. In all, CSM presents a number of advantages over DSM, and is applicable to more practical scenarios.